# On dielectric screening in twisted double bilayer graphene

Fumiya Mukai[1], Kota Horii[1], Nazuna Hata[1], Ryoya Ebisuoka[2,] Kenji Watanabe[3], Takashi Taniguchi[4] and Ryuta Yagi[1,2*]

[1)]Graduate School of Advanced Science and Engineering, Hiroshima University, Kagamiyama 1-3-1, Higashi-Hiroshima, Hiroshima 739-8530, Japan,
[2)] Graduate School of Advanced Sciences of Matter, Hiroshima University, Kagamiyama 1-3-1, Higashi-Hiroshima, Hiroshima 739-8530, Japan,
[3)]Research Center for Functional Materials, National Institute for Materials Science, 1-1 Namiki, Tsukuba 305-0044, Japan,
[4)]International Center for Materials Nanoarchitectonics, National Institute for Materials Science, 1-1 Namiki, Tsukuba 305-0044, Japan.

ABSTRACT

We have studied the dielectric screening of electric field which is induced by a gate voltage in twisted double bilayer graphene by using a sample with a mismatch angle of about 5 degrees. In low temperature magnetotransport measurements, quantum oscillations of magnetoresistance originating from two bands with different carrier density were observed. The behavior of the carrier densities with respect to the total carrier density were distinct from that of the AB-stacked tetralayer graphene. The carrier density ratio was theoretically analyzed in terms of the model that the induced charge decays exponentially with distance with a screening length $\lambda$. The estimated $\lambda$ was slightly larger than that of AB-stacked graphene, which would possibly reflect the difference in the inter-plane distribution of probability of the wave function.

*Corresponding author, yagi@hiroshima-u.ac.jp

## 1. Introduction

Graphene has been studied by a numerous research activities since its discovery, in terms of its fundamental physics as well as its potential application. Multilayer graphene has two stable stacking structures, ABA or ABC, in terms of relative position of the carbon atoms between neighboring layers. It is known that the electronic band structure of multilayer graphene is sensitive to the stacking structure; ABA-stacked graphene shows the even-odd layer number effect on the regularity of numbers of massive and massless band [1-22]; ABC-stacked graphene shows a dispersion relation obeying the power law, whose power depends on the number of layers. Owing to the development of the technique to stack atomic layers in a controlled manner, it is now possible to fabricate twisted graphene and other similar systems with an arbitrary twist angle between the crystallographic axes [23-25]. In case of the stack of atomic layers with approximately the same structure, for example, hexagonal boron nitride ($h$-BN) and graphene, a small mismatch angle of the crystallographic axes causes a geometrical resonance (moiré) of the top and bottom layer. The atomic moiré acts as an effective potential through the variation of the coupling between the top and the bottom layer, and the system becomes a superlattice. In a transport property, there appear satellite resistance peaks arising from forming secondary Dirac cones created by the band folding associated with the superlattice potential [26-31]. Twisted bilayer graphene, a stack of two graphene sheets, also exhibits particular phenomena arising from the moiré structure [25,32-49]. A flat band appears for the mismatch angle of about 1.1°[43-45], and correlated insulating states and even superconductivity [50] were found near the angle. Twisted double bilayer graphene, a twisted stack of bilayer graphene, has also been studied [51-56]. The dispersion relation can be widely tuned by applying the perpendicular electric field [51,57-58]. The twisted double bilayer graphene also shows the similar insulating states and a superconductivity-like behavior [52-54,56].

While the studies of twisted double bilayer graphene focus on the cases around the magic angle, the cases with large mismatch angles has been less explored. Because the mismatch angle is not close to the flat band condition, low-energy band structure of the whole system can be roughly approximated by independent bands of the two bilayer graphene flakes. Hybridization of bands in the low energy regime is virtually negligible because the two bands are separated in the $k$-space. Therefore, it is naturally expected that the wave function in the direction perpendicular to the two-dimensional plane is significantly different from that of AB stacked tetralayer graphene. In the former case, wave function would approximately localize either top or bottom bilayer graphene layer, while, in the latter, it extends over the layers.

In this paper, we studied the dielectric screening in the twisted double bilayer graphene. Gate voltages alter not only carrier density, but also electrostatic potentials in the graphene layers, which is closely related with a dielectric screening. Studies on dielectric screening in AB-stacked graphene can be sought in the similar ones in the graphene intercalation compounds. It was theoretically predicted that the Thomas Fermi screening length is about 0.38 - 0.5 nm (1-2 layer) if there are no interlayer tunneling [59-61]. Guinea studied the screening in presence of interlayer hopping [62]. Koshino studied dielectric screening by self-consistent calculation of the wave function and potential for each layer [63]. He found that the behavior of the dielectric screening reflects the wave function in the direction perpendicular to two-dimensional plane. Experimentally, there are some efforts to determine the screening length. Ohta *et al.* estimated $\lambda$ to be 0.14 - 0.19 nm by angle resolved photoemission spectroscopy (ARPES) measurement [64]. Miyazaki modelled the system as a stack of independent graphene sheets, and estimated $\lambda$ to be $3-7$ by using transport measurements of multilayer graphene [65]. The present authors estimated $\lambda$ to be about 1 layer from the detailed comparison of theory and experiment on the Landau level structure and intrinsic resistance peaks [18,20].

## 2. Experimental

Twisted double bilayer graphene sample was made from a bilayer graphene flake prepared by the mechanical exfoliation of Kish graphite. To form the twisted structure, the tear-and-stack method was used [25]. Fig. 1(a) shows an illustration to form twisted double bilayer graphene. Fig. 1(b) and 1(c) show the vertical structure and an optical micrograph of our sample. The sample was encapsulated with thin *h*-BN flakes to protect surface of graphene from contamination [23], and to form the gate electrodes. Between the encapsulated graphene and gate electrodes, there were relatively thick *h*-BN flakes. These reduced leakage current from the gate electrodes to the effective sample area, and also reduced a probability of electric breakdown which might be caused by applying large gate voltages. The sample was patterned into a Hall bar by using standard electron beam lithography and reactive ion etching with a mixture of low pressure $O_2$ and $CF_4$ gas. The electronic contacts with the graphene sample were attained by using the edge-contact technique [23]. The angle between the crystallographic axes of the top and the bottom layer was estimated to be about 5 degrees by using optical micrographs. Resistivity measurement was performed by using standard lock-in technique. The magnetic field was applied with a superconducting solenoid.

3. Result and discussion

Fig. 1(d) shows resistivity ($\rho$) vs gate voltage ($V_b$), which was measured at the helium temperature ($T$ = 4.2 K). The sample's electronic mobility was estimated to be about 3000 cm²/Vs for large carrier densities ($n_{tot}$). (The carrier density was determined from the gate voltage by using a result of quantum oscillation measurements which will be shown later.) Twisted double bilayer graphene also shows satellite resistance peaks due to secondary Dirac points formed as a result of the folding of the band structure. Our sample does not show the peaks in the measured range of carrier density. The peaks are expected to appear at much larger carrier densities above $1 \times 10^{13}$ cm$^{-2}$ because sample's mismatch angle is about 5 degrees.

Low temperature ($T$ = 4.2 K) magnetoresistance exhibited Shubnikov-de Haas effect. Fig. 2(a) is a map of the longitudinal resistivity ($\rho_{xx}$), plotted as a function of carrier density ($n_{tot}$) and magnetic field ($B$). Each stripe originates from a Landau level. The structure is more clearly visible in Fig 2(b), which is a map of $d\rho_{xx}/dB$. The Landau fan diagram is different from those of the any other AB-stacked ones [18]. This indicates that the present system is distinct from neither AB-stacked tetralayer graphene nor bilayer graphene. Indeed, only a single zero-mode Landau level was visible at the charge neutrality point. This contrasts with the case of AB-stacked tetralayer graphene; the zero-mode Landau levels of the heavy- and the light-mass bilayer-like band appear at different carrier densities [17-19]. The zero-mode Landau level around the charge neutrality point has 16-fold degeneracy, and is the twice the degeneracy of that of the zero-mode Landau level of the bilayer graphene. On the other hand, higher Landau levels show complicated crossings, which are absent from pristine bilayer graphene. Therefore, the Landau level structure in Figs. 2(a) and 2(b) reflected the particular electronic band structure of twisted double bilayer graphene.

Detailed band structure manifested itself by finding the periodicity in the $\rho_{xx}$ vs $1/B$ traces, from which one can calculate the areas of the Fermi surface resulting in the S-dH oscillations. The area is related with the carrier density of the band. Fig. 2 (c) shows a map of the power spectrum of the fast Fourier transformation (FFT). A FFT frequency was converted into carrier densities (for bands with four-fold degeneracy due to a spin and a valley degree of freedom). In the figure, one can see two Fourier components (band1 and band2) for both the electron and the hole regime. Power spectrum of the Fourier components in the electron regime appeared larger than that in the hole regime because the S-dH oscillation amplitude is larger in the electron regime than that in the hole regime. The low-energy band of the twisted double bilayer graphene consists of band1 and band2 because the sum of the two carrier density equals total carrier density induced by the gate voltage. The carrier density of band1 is about a factor of 2.5 (2.7) larger than that of band2 in the electron (hole) regime. This ensures that band1 did not originate from higher order harmonics of band2, but stemmed from a different band because the carrier density (or the FFT frequency) of band1 was not an integer multiple of that of band2.

In the absence of perpendicular electric field, the bilayer graphene flakes at the top and the bottom layer of the twisted double bilayer graphene are energetically equivalent so the bands are degenerated. Perpendicular electric field breaks the symmetry of the top and the bottom bilayer graphene. Similar feature was also seen in in twisted bilayer graphene with not small mismatch angle [66]. In the present experiment, to change the carrier density, a perpendicular electric field is inevitably applied by the gate voltage of single gate electrode [18]. In the graphene with multiple layers, the band structure is strongly influenced by the externally applied perpendicular electric field, which is screened layer by layer in the graphene sample. Textbooks on electromagnetism tell that the induced charges distribute only at the surface of the conductor, and no electric field is present inside the conductor. This is not correct for graphene and other atomically thin materials. To date, the screening of external electric field has been studied theoretically [59-63] and experimentally [18,64-65]. First of all, we analyze our experimental result by using the simplest model similar to the Thomas-Fermi approximation, in which gate-voltage induces charges in each layer with an exponential attenuation with distance (or layer numbers). The charge $n_j$ for $j$-th layer from the gate electrode is

$$n_j = n_{tot} \exp(-\frac{jd}{\lambda}) / \sum_{j=1}^{4} \exp\left(-\frac{jd}{\lambda}\right), \tag{1}$$

where $d$ is interlayer distance, $\lambda$ is a screening length. In this simple model, sum of carriers in layer 1 and 2 ($n_1 + n_2$) is

$$n_1 + n_2 = n_{tot}\left(\exp\left(-\frac{d}{\lambda}\right) + \exp\left(-\frac{2d}{\lambda}\right)\right) / \sum_{j=1}^{4} \exp\left(-\frac{jd}{\lambda}\right), \tag{2}$$

and that in layer 3 and 4 ($n_3 + n_4$) is given by

$$n_3 + n_4 = n_{tot}(\exp(-\frac{3d}{\lambda}) + \exp(-\frac{4d}{\lambda})) / \sum_{j=1}^{4}\exp(-\frac{jd}{\lambda}) = \exp\left(-\frac{2d}{\lambda}\right)(n_1 + n_2). \tag{3}$$

Using present experimental values, $n_{12}/n_{34} \sim$ 2.5 in electron regime and $n_{12}/n_{34} \sim$2.7 in hole regime, one could estimate $\lambda = 0.68 - 0.74$ nm, *i.e.*, about 2 layers. This value is approximately the same as the theoretical values of the graphite intercalation compounds [59-61] and the values obtained in the early experiments on graphene [64-65]. However, it is considerably larger than the value ($\lambda \sim 0.35$ nm) of the recent experimental and theoretical works on the AB-stacked multilayer graphene [18].

A more elaborate method to estimate $\lambda$ is to calculate the band structure. Figure 3 (a) shows a schematic superlattice Brillouin zone in the twisted double bilayer graphene. The superlattice Brillouin zone is a hexagon, whose adjacent vertices are K (K') point of the top and the bottom bilayer graphene, which are rotated by mismatch angle $\theta$. The electronic band structure of twisted double bilayer graphene was calculated by considering the electrostatic potentials for each layer, which is induced by the gate voltage [18,20,31]. (See Appendix.) If one ignores stray capacitance, electric field negligibly escapes through the surface of the graphene, which is located at the other side of the gate electrode. Then, one can use formulae,

$$n_{tot} = C_b V_b / e, \tag{5}$$

and

$$D_\perp = C_b V_b / 2, \tag{6}$$

where $D_\perp$ is the perpendicular electric flux density, $C_b$ is the bottom gate capacitance, $V_b$ is the bottom gate voltage. Therefore, the perpendicular electric field varies with $V_b$. For a fixed carrier density, carrier distribution in each layer was calculated by using eq. (1) with $\lambda$ as an adjustable parameter to be determined. Electrostatic potential was calculated by using the Poisson equation with relative permittivity, $\epsilon/\epsilon_0 = 2$. The potential was added in the diagonal element of the Hamiltonian of the continuum model. The effective moiré potential for the twisted bilayer graphene was also considered in the second and third layer [51].

First, we show a result for the case without considering perpendicular electric field. Fig. 3(b) shows Fermi surfaces for $|n_{tot}| = 3.0 \times 10^{12}$ cm$^{-2}$. Because top and bottom bilayer graphene are energetically equivalent, their Fermi surfaces have the same cross-sectional areas (or carrier densities) and appear at $\bar{K}$ and $\bar{K}'$ point (Fig. 3(c)). Each of these stems from a Fermi surface of the original bilayer graphene at the top or the bottom layer of the stack. Apparently, this result infers that one must take into account the electrostatic potential arising from the gate voltage. Fig. 3(d) shows the result for a condition which approximately reproduced the experimentally observed ratio of the Fermi surfaces. The reason why the areas of the two Fermi surfaces at $\bar{K}$ and $\bar{K}'$ point are different is simple. The top and the bottom bilayer graphene are approximately independent, so that the offset energy of the bands approximately reflects the mean electrostatic potential of each layer. Ratio of the carrier density of the smaller Fermi surface ($n_s$) to $n_{tot}$ varies with $\lambda$ as shown in Figs. 4(a) and 4(b). By using experimental values, $n_s/n_{tot} \approx 0.29$ in the electron regime and $n_s/n_{tot} \approx 0.27$ in the hole regime, one could estimate $\lambda \sim 0.47$ nm in the electron regime, and $\lambda \sim 0.42$ nm in the hole regime. Note that the optimum screening length slightly depended on the Slonczewski-Weiss-McClure (SWMcC) parameters. If one uses parameters ($\gamma_0 = 2.47$ eV, $\gamma_1 = 0.4$ eV) used in Ref. [51] $\lambda = 0.3 - 0.35$ nm.

The effect of perpendicular electric field, resulting from applying a gate voltage by using single gate electrode, has also been observed in AB-stacked multilayer graphene [18]. The AB-stacked tetralayer graphene is a good example to compare with the present case because of the same number of layers. The AB-stacked tetralayer graphene has two bands, the light-mass and the heavy-mass bilayer-like band. The result of the experiment could not be explained unless gate-voltage, induced by the perpendicular electric field, was considered. Figs. 4(c) and 4(f) show results of numerical calculation on $n_s/n_{tot}$ vs $\lambda$ for SWMcC parameter of graphite. In the previous experiment (Fig. 4 in Ref. [18]), $n_s/n_{tot} \approx 0.20$ in the electron regime and $n_s/n_{tot} \approx 0.27$ in the hole regime at $|n_{tot}| = 3.0 \times 10^{12}$ cm$^{-2}$. Using Figs. 4(c) and 4(f), $\lambda \sim 0.31 - 0.32$ nm consistently explained experiment. Here, $\lambda$ is slightly dependent on the choice of SWMcC parameters.

Screening length of twisted double bilayer graphene is the same order of magnitude as that of the AB stacked teteralayer graphene, but it is possibly larger than that of AB-stacked tetralayer graphene, which would reflect the different eigenstates in the stacking direction.

We also performed self-consistent calculation of $n_s/n_{tot}$. In the calculation, eigenvectors were considered to calculate the carrier distribution in each layer. The calculation was done until the carrier distribution is self-consistently determined. In the calculation, the SWMcC parameters of graphite were used, and $\epsilon/\epsilon_0$ was varied as an adjustable parameter. Results of $n_s/n_{tot}$ for $|n_{tot}| = 3 \times 10^{12} \mathrm{cm}^{-2}$ is shown in Fig. 4(e) and 4(f). The experiment was reproduced when $\epsilon/\epsilon_0 \approx 3.5$, which is slightly larger than that of graphite. The reason of the enhancement of $\epsilon/\epsilon_0$ is currently unclear. Although choice of SWMcC parameters might be concerned, one cannot deny a possibility of many body effects. This issue needs further investigation.

## 4. Conclusion

Dielectric screening of the electric charges, which are induced by the gate voltage, was studied in a twisted double bilayer graphene sample with a mismatch angle of 5 degrees. By applying a gate voltage only on one side of the twisted double bilayer graphene, we studied screening effect by observing Shubnikov-de Haas oscillations. The oscillations had two different periods, which resulted from different electrostatic potentials between top and bottom bilayer graphene. Carrier densities for top bilayer graphene to the bottom bilayer graphene was about factor of 2.5-2.7 larger than that of the bottom layer, with which screening length was estimated to be 1-2 layers. Estimation of $\lambda$ by numerical calculation of the band structure was also done. The estimated $\lambda$ was slightly larger in twisted double bilayer graphene than in AB-stacked tetralayer graphene. This might reflect the difference of the band structure between twisted double bilayer graphene and AB-stacked graphene.

# Appendix A

## Method of band calculation

Band structure of twisted double bilayer graphene was calculated by using the Hamiltonian based on effective mass approximation [51],

$$H = \begin{pmatrix} H_0(\boldsymbol{k_1}) + \Phi_1 & g^\dagger(\boldsymbol{k_1}) & 0 & 0 \\ g(\boldsymbol{k_1}) & H_0'(\boldsymbol{k_1}) + \Phi_2 & U^\dagger & 0 \\ 0 & U & H_0(\boldsymbol{k_2}) + \Phi_3 & g^\dagger(\boldsymbol{k_2}) \\ 0 & 0 & g(\boldsymbol{k_2}) & H_0'(\boldsymbol{k_2}) + \Phi_4 \end{pmatrix}. \quad (7)$$

Here,

$$H_0(\boldsymbol{k}) = \begin{pmatrix} 0 & -\hbar v k_- \\ -\hbar v k_+ & \Delta' \end{pmatrix}, \quad (8)$$

$$H_0'(\boldsymbol{k}) = \begin{pmatrix} 0 & -\hbar v k_- \\ -\hbar v k_+ & \Delta' \end{pmatrix}, \quad (9)$$

$$g(\boldsymbol{k}) = \begin{pmatrix} \hbar v_4 k_- & \gamma_1 \\ \hbar v_3 k_+ & \hbar v_4 k_- \end{pmatrix}, \quad (10)$$

and

$$\Phi_i = \begin{pmatrix} 1 & 0 \\ 0 & 1 \end{pmatrix} \phi_i, \quad (11)$$

where $k_\pm = -i\xi \frac{\partial}{\partial k_x} \mp \frac{\partial}{\partial k_y}$, $\xi\ (=\pm 1)$ is a valley index, and $\phi_j$ (j = 1 − 4) is an electrostatic potential of $j$-th layer. We used the effective moiré potential which was derived by Koshino [51],

$$U = \begin{pmatrix} u & u' \\ u' & u \end{pmatrix} + \begin{pmatrix} u & u'\omega^{-\xi} \\ u'\omega^{\xi} & u \end{pmatrix} \exp(i\xi \boldsymbol{G_1^M} \cdot \boldsymbol{r}) + \begin{pmatrix} u & u'\omega^{\xi} \\ u'\omega^{-\xi} & u \end{pmatrix} \exp(i\xi(\boldsymbol{G_1^M} + \boldsymbol{G_2^M}) \cdot \boldsymbol{r}). \quad (11)$$

Here, $\omega = \exp\left(\frac{2\pi i}{3}\right)$, $u = 0.0797\ eV$, $u' = 0.0975\ eV$, $v = \sqrt{3}a\gamma_0/(2\hbar)$, $\xi = \pm 1$ is a valley index, and $a$ is a lattice constant of graphene. $\boldsymbol{G_1^M}$ and $\boldsymbol{G_2^M}$ are reciprocal vectors of the superlattice Brillouin zone. We calculated energy eigenvalues numerically by expanding the wave function into plane waves.

Dependence on parameters

Profiles of $n_s$ vs $n_{tot}$ is slightly dependent on the SWMcC parameters which was used in the calculation. Fig. 5(a) and 5(b) show the results with $\gamma_0 = 2.466$ eV, and $\gamma_1 = 0.4$ eV. The experimentally observed $n_s/n_{tot}$ is approximately reproduced for $\lambda \sim 0.31 - 0.35$ nm. The optimum value of $\lambda$ is slightly smaller than those of the calculation with SWMcC parameter of graphite, however, it remains about 1 layer. We also studied the dependence on the potential intensity. Fig. 5(c) and 5(d) show the similar results for half $u$ and $u'$. The experimental values were reproduced for $\lambda \sim 0.31 - 0.38$ nm. The magnitude of the moiré potential does not influence the relation of $n_s/n_{tot}$ vs $\lambda$ significantly.

# Figure Caption

Fig. 1

(a) A schematic illustration of formation of a twisted double bilayer graphene sample. Two flakes of bilayer graphene, which were split from a flake, were stacked on a $h$-BN flake. Relative angle $\theta$ between the crystallographic axes of bilayer graphene was 5°. hBN is $h$-BN, BLG is bilayer graphene, and TDBLG is twisted double bilayer graphene. (b) A schematic vertical structure of the sample. Twisted double bilayer graphene is encapsulated with $h$-BN. It was sandwiched by a pair of graphite gate electrodes; thick $h$-BN flakes were inserted between them for insulation. Top gate was accidentally shorted to ground so that the carrier density was tuned only by applying gate voltage to bottom gate electrode. GGate is a graphite gate electrode. (c) Optical micrograph of our device. (d) $\rho$ vs $V_b$ measured at $T = 4.2$ K. (e) Hall mobility $\mu$ vs carrier density $n_{tot}$. ($n_{tot}$ was calibrated by using the Shubnikov de Haas effect). $T = 4.2$ K.

Fig. 2

(a) A map of $\rho_{xx}$ as a function of carrier density $n_{tot}$ and magnetic field $B$. $T = 4.2$ K. (b) A similar map for $d\rho_{xx}/dB$. In panels (a) and (b), the filling factor $\nu$ is displayed with numbers and dashed lines. (c) A map of Fourier spectra. The horizontal axis is the total carrier density ($n_{tot}$) that was induced by the gate voltage. The vertical axis is the carrier density of a Fermi surface, which is proportional to the frequency of the Shubnikov-de Haas oscillation that is periodic in $1/B$. Carrier density was calculated by taking into account of four-fold degeneracy due to spin and valley of the original bilayer graphene flakes which consisted the sample.

Fig. 3

(a) Schematic drawing of the moiré Brillouin zone of a twisted double bilayer graphene. $\bar{K}$ and $\bar{K}'$ point are corners of the moiré Brillouin zone. (b) Numerically calculated energy contour of the dispersion relation for twisted double bilayer graphene in the absence of the electrostatic potential induced by the gate voltage. A pair of trigonally warped Fermi surfaces, appears at both $\bar{K}$ and $\bar{K}'$ point, which are identical in shape. The fermi surface at $\bar{K}$ ($\bar{K}'$) originates from the top (bottom) bilayer graphene. Cases for $|n_{tot}| = 3 \times 10^{12}$ cm$^{-2}$ are shown. Only the Fermi surfaces arising from K point is plotted. (c) Schematic illustration of the dispersion relation along $\bar{K}$ - $\bar{K}'$ line. (d) An example of numerically calculated fermi surface for $\lambda = 0.50$ nm (left; electron regime) and 0.4 nm (right; hole regime). Cases for $|n_{tot}| = 3 \times 10^{12}$ cm$^{-2}$ are shown.

Only the Fermi surfaces arising from K point is plotted. (e) Schematic illustration of the dispersion relation along $\bar{K}$ - $\bar{K}'$ line. Energy gaps open at the bottoms of the bilayer bands. Because of forming the energy gap, and difference in the electrostatic potential, the energy of the bottoms of the bands offset with different values. For panels (b) and (d), SWMcC parameters of graphite ($\gamma_0 = 3.16$ eV, $\gamma_1 = 0.39$ eV, $\gamma_3 = 0.3$ eV, $\gamma_4 = 0.044$ eV, and $\Delta' = 0.037$ eV) were used.

Fig. 4
Theoretical $n_s/n_{tot}$ which was calculated for $|n_{tot}| = 3 \times 10^{12}$ cm$^{-2}$. Panels (a) and (b) show the results in twisted double bilayer graphene with the screening model with a relaxation length $\lambda$. Panels (c) and (d) show similar results for AB-stacked tetralayer graphene. Panels (e) and (f) show self-consistent calculations for different values of $\epsilon/\epsilon_0$. SWMcC parameters of graphite were used in all the calculations. Broken likes indicates experimental value of $n_s/n_{tot}$, and $\lambda$ estimated by the interpolation of the numerical calculations. TDB and AB denote twisted double bilayer graphene, and AB-stacked tetralayer graphene, respectively

Fig. 5
Theoretical calculation of $n_s/n_{tot}$ for $|n_{tot}| = 3 \times 10^{12}$ cm$^{-2}$. $\gamma_0 = 2.466$ eV, $\gamma_1 = 0.39$ eV, $\gamma_3 = 0.3$ eV, $\gamma_4 = 0.044$ eV, and $\Delta' = 0.037$ eV. Panels (a) and (b) show the results for electron and hole regime, which were calculated by using $u = 0.0797$ eV and $u' = 0.0975$ eV. Panels (c) and (d) are similar results for half $u$ and $u'$.

Fig. 1

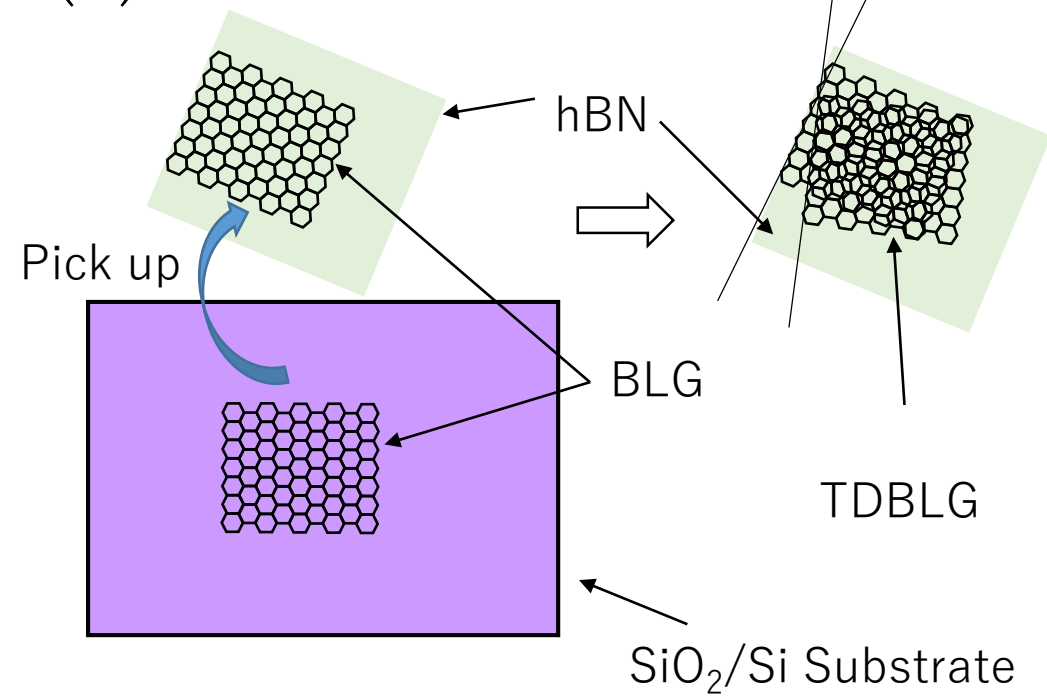
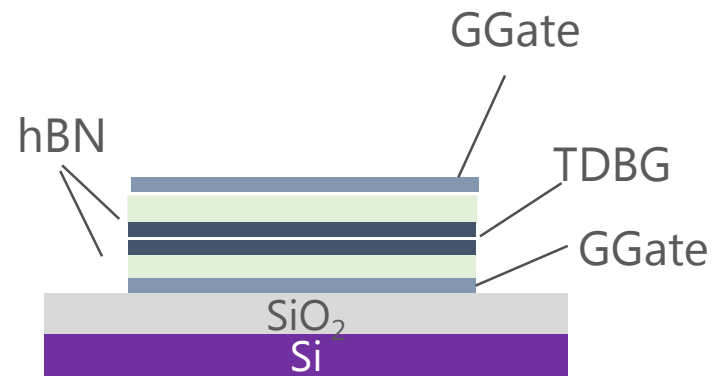
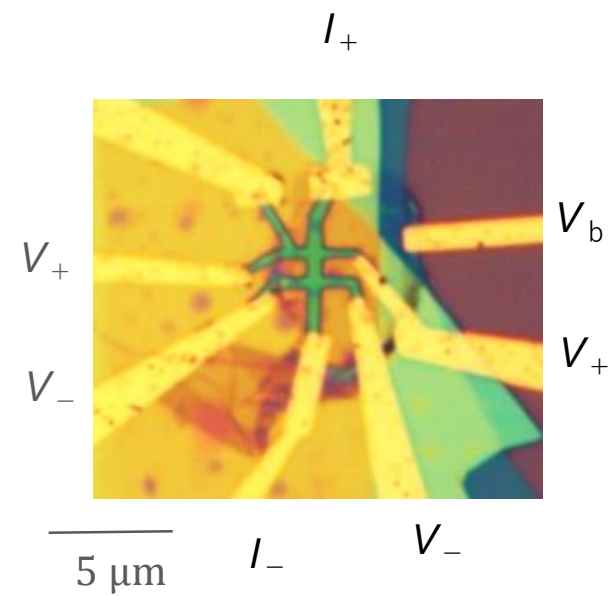
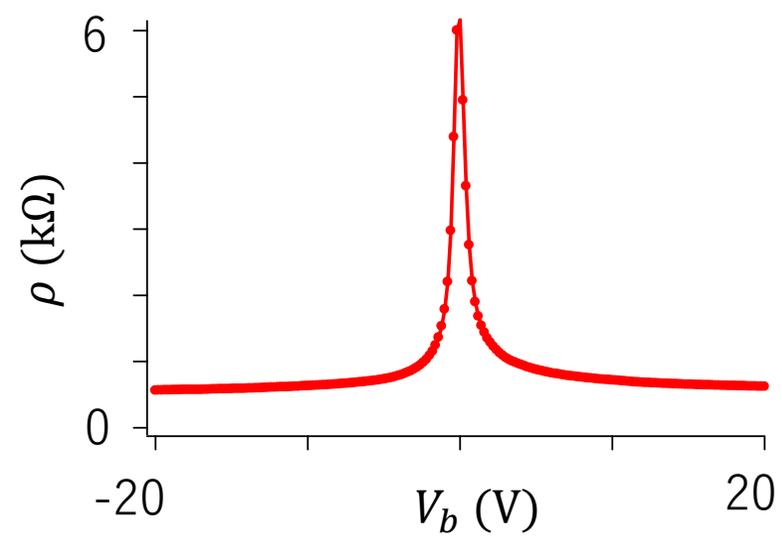
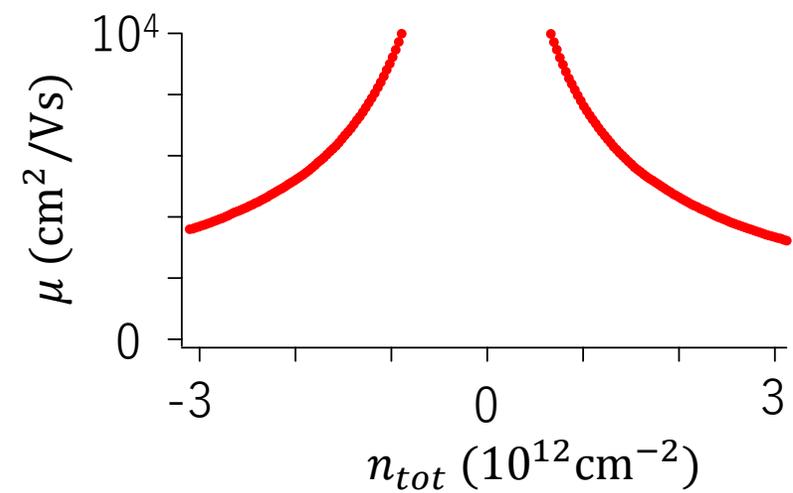

Fig. 2

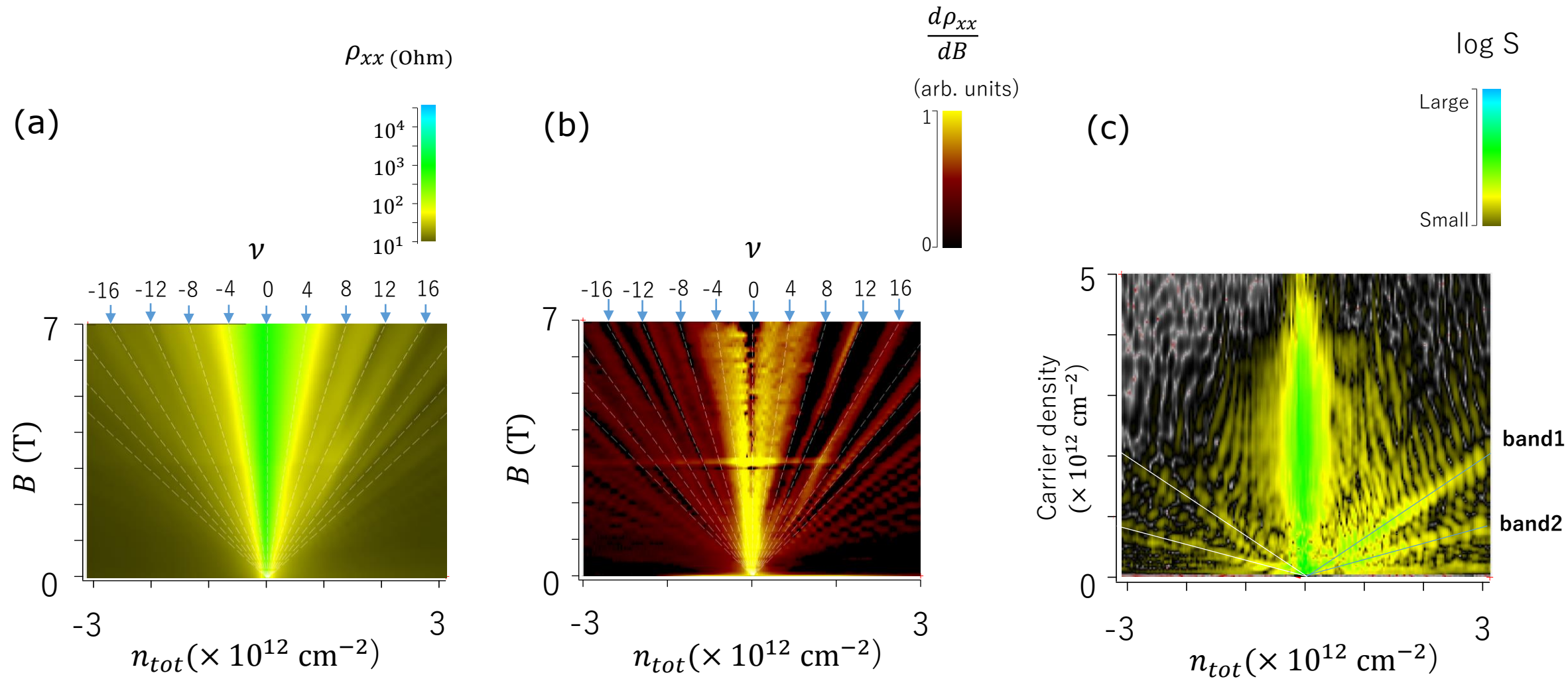

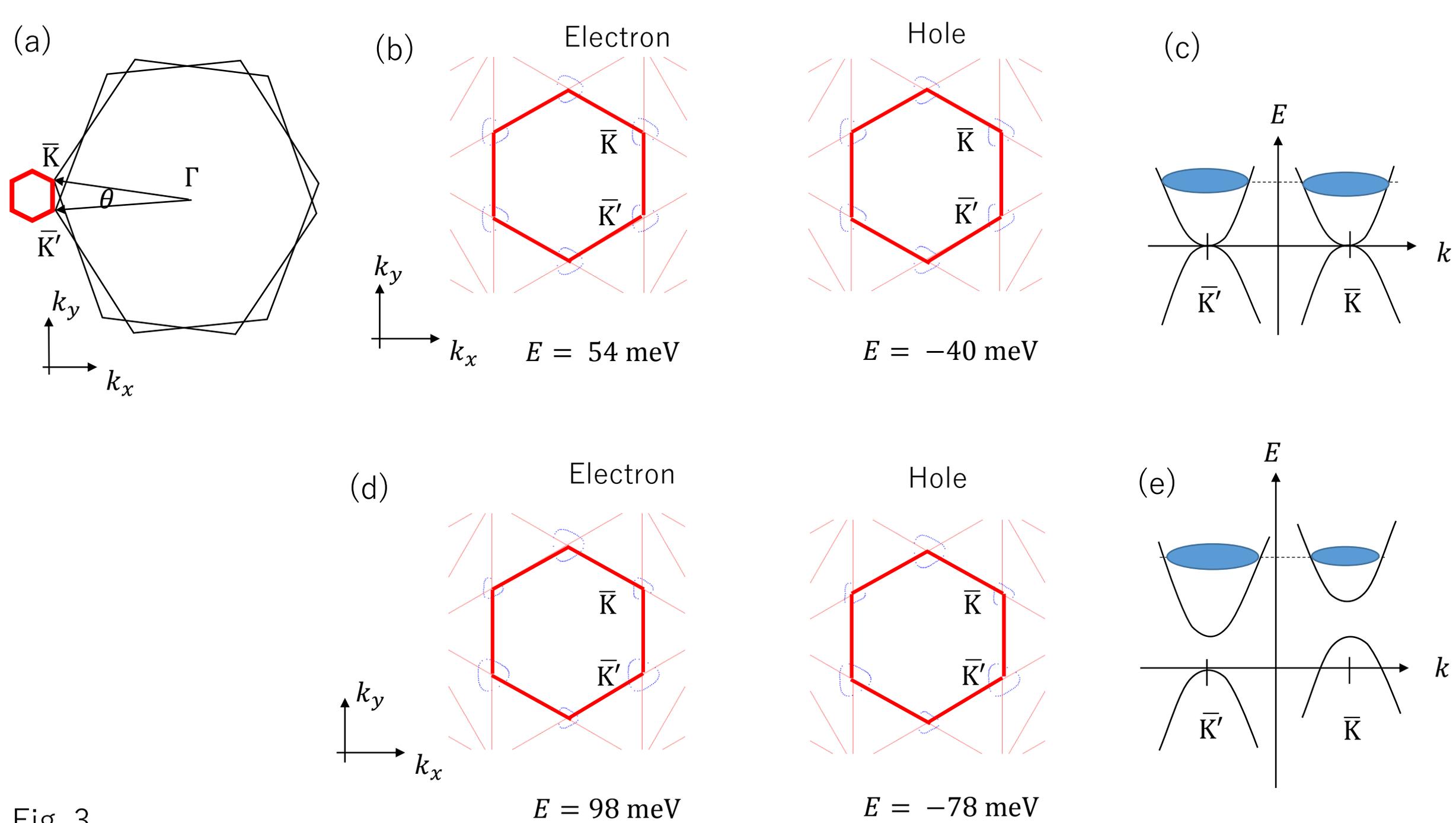

Fig. 3

Fig. 4

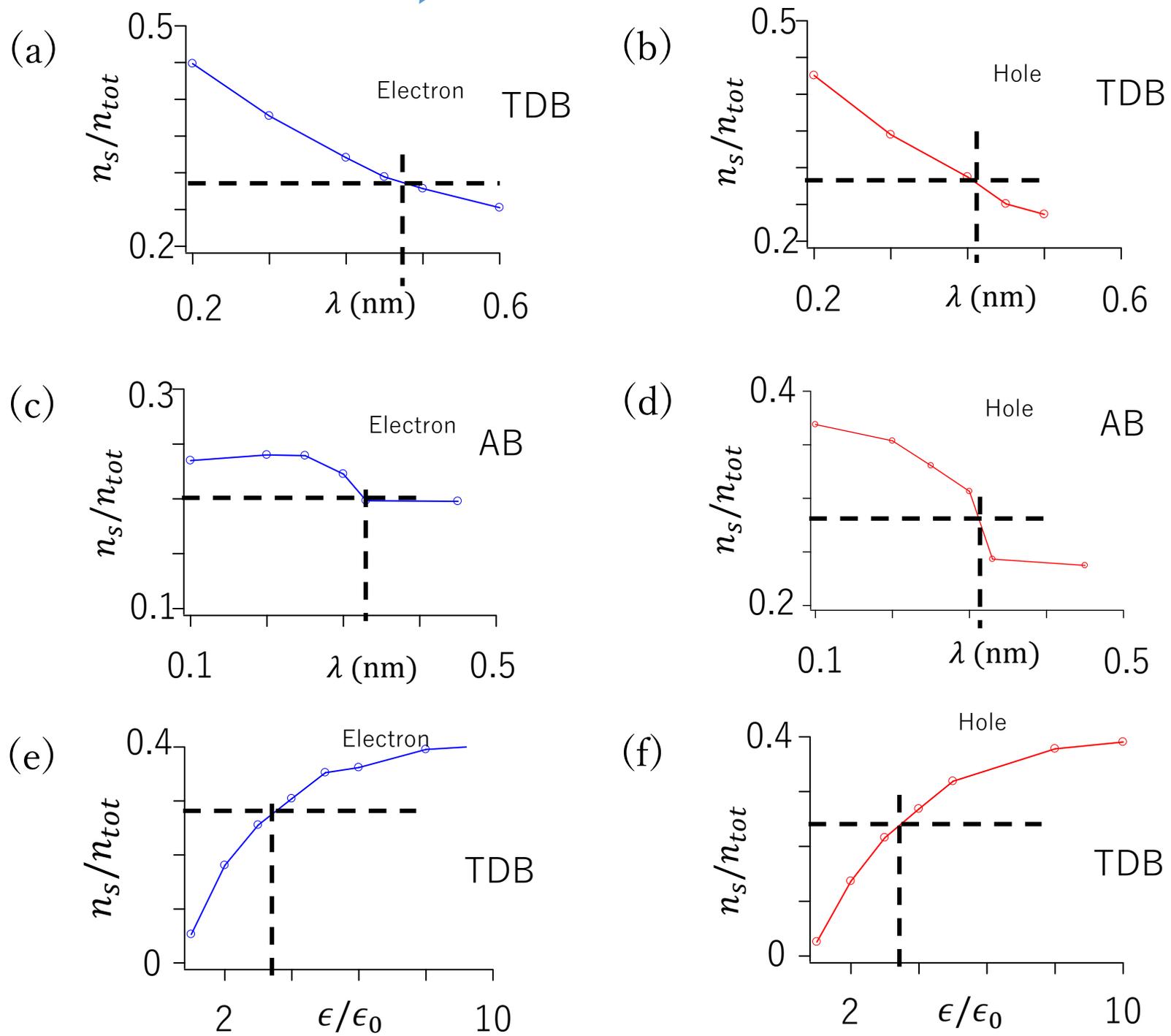

Fig. 5

(a)
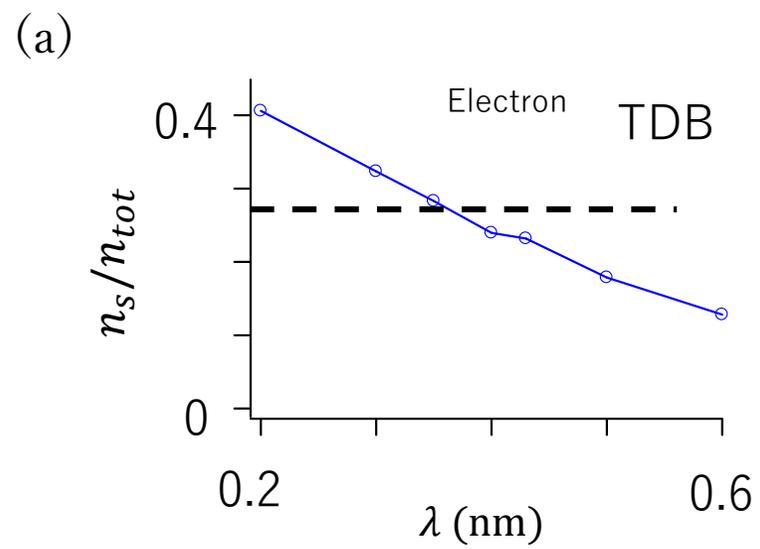

(b)
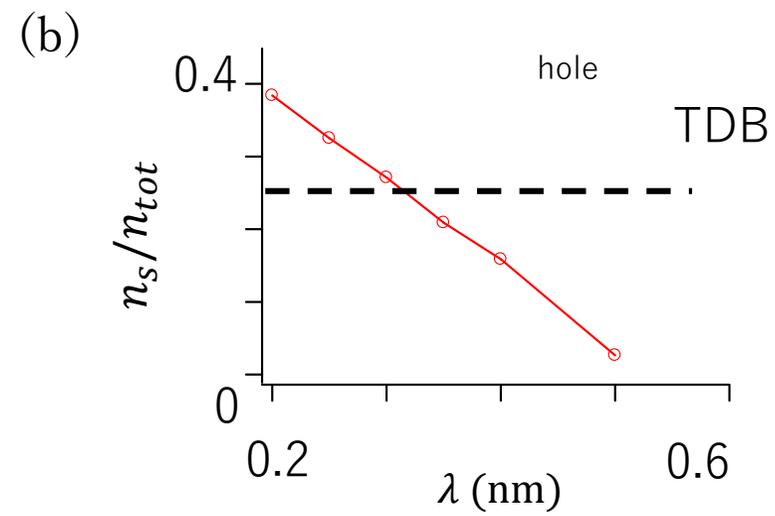

(c)
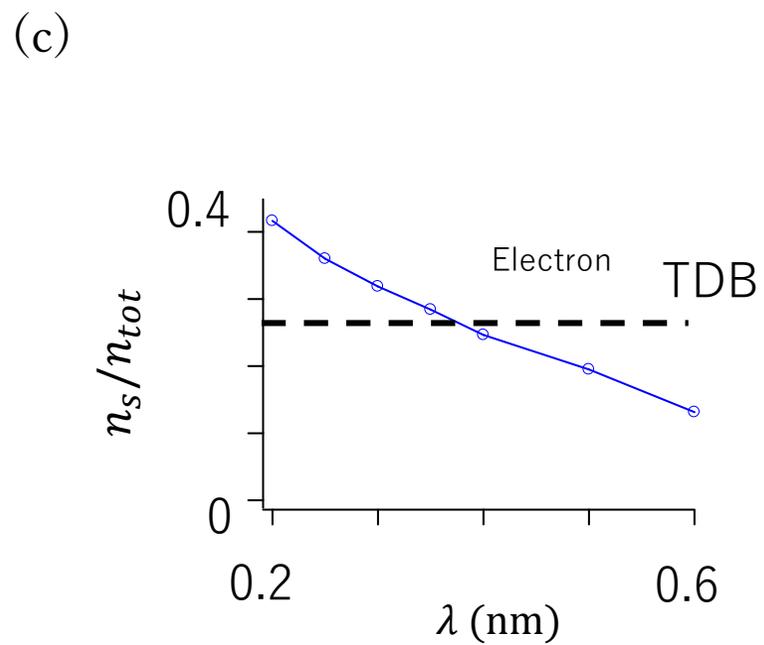

(d)
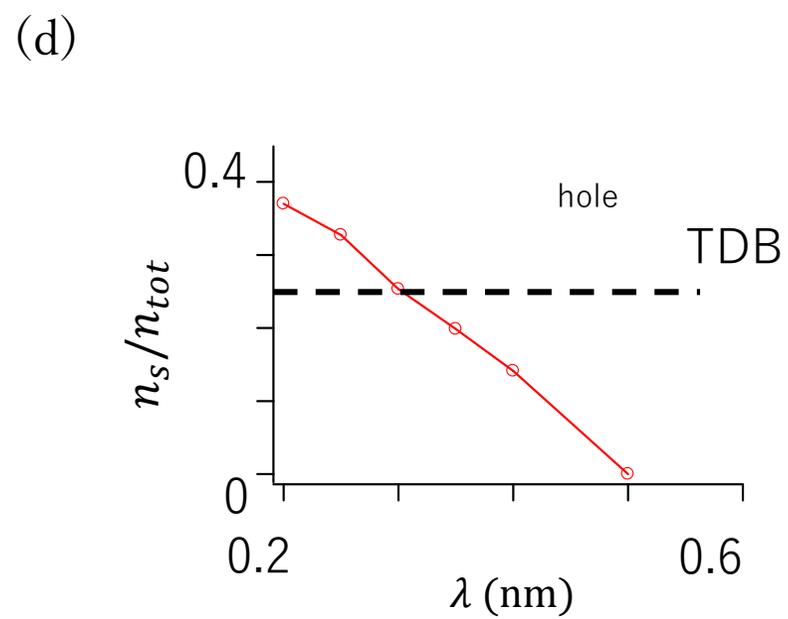